\documentclass[aps,twocolumn,groupedaddress,prb,floatfix,showpacs]{revtex4}
\usepackage{graphicx}
\usepackage{dcolumn}
\usepackage{bm}
\usepackage{amsmath}
\usepackage{verbatim}
\begin{document}

\title{Effects of magnetism and electric field on the energy gap of bilayer graphene nanoflakes}

\author{Bhagawan Sahu$^1$}
\email{brsahu@physics.utexas.edu}
\author{Hongki Min$^{2,3}$}
\author{Sanjay K. Banerjee$^1$}
\affiliation{
$^1$Microelectronics Research Center, The University of Texas at Austin, Austin, TX 78758\\
$^2$Center for Nanoscale Science and Technology, National Institute of Standards and Technology, Gaithersburg, MD 20899-6202 \\
$^3$Maryland NanoCenter, University of Maryland, College Park, MD 20742
}

\date{\today}

\begin{abstract}
We study the effect of magnetism and perpendicular external electric field strengths on the energy gap of length confined bilayer graphene nanoribbons (or nanoflakes) 
as a function of ribbon width and length 
using a \textit{first principles} density functional electronic structure method and a semi-local exchange-correlation approximation. We assume AB (Bernal) bilayer stacking and consider both armchair and zigzag edges, and for each edge type, we consider the two edge alignments, namely, $\alpha$ and $\beta$ edge alignment. For the armchair nanoflakes we identify three distinct classes of bilayer energy gaps, determined by the number of carbon chains in the width direction ({\it N} = 3{\it p}, 3{\it p}+1 and 3{\it p}+2, {\it p} is an integer), and the gaps decrease with increasing width 
except for class 3{\it p}+2 armchair nanoribbons. 
Metallic-like behavior seen in armchair bilayer nanoribbons are found to be absent in armchair nanoflakes. Class 3{\it p}+2 armchair nanoflakes show significant length dependence. 
We find that the gaps decrease with the applied electric fields due to large intrinsic gap of the nanoflake. The existence of a critical gap with respect to the applied field, therefore, is not predicted by our calculations.
Magnetism between the layers plays a major role in enhancing the gap values resulting from the geometrical confinement, hinting at
an interplay of magnetism and geometrical confinement
in finite size bilayer graphene. 
\end{abstract}

\maketitle

\section{Introduction}

Bilayer graphene continues to attract attention from the condensed matter physics community as well as from  industry because of the prediction and subsequent experimental realization of electric field opening of the band gap,\cite{sahu} the theoretical predictions of a novel electronic phase and the theoretical realization of a novel charge based electronic switch.\cite{hongki} 
The modulation of bilayer gaps by a geometrical confinement (in particular by widths), interlayer magnetism and edge alignments\cite{sahu1, lima} offers exciting opportunities for designing novel electronic and optical devices.  

Recent advancement in the growth of large area graphene films on metal substrates,\cite{rod} deposition of a dielectric layer on the graphene film,\cite{tutuc} and  making contacts to graphene for transport studies\cite{goldhaber} motivates
the theoretical understanding of the intrinsic (edges and their alignments, edge disorders, layer stackings) as well as extrinsic (substrates, contacts, dielectrics, adatoms, defects) perturbations to the ideal properties of graphene and graphene sheets. 
In transport measurements, the edges of the graphene flakes which are in electrical contact with a piece of metal or an another piece of graphene play a decisive role in stabilizing an interface potential barrier for flow of electrons across it and this potential barrier depends on the {\it intrinsic} gap of the graphene flake. 

Theoretical studies of 
isolated finite size graphene flakes are, therefore, critical to understand the tunability of the gap value resulting from the edge confinements.  
Moreover, magnetism which is absent in bulk graphene bilayer can appear because of edges in the finite size graphene flakes. 
In addition, in bilayer systems, one can apply an external electric field to tune the energy band gap.\cite{sahu} 
It is, therefore, desirable to understand the effects of magnetism and external electric fields on the energy gaps of geometrically confined isolated graphene sheets.

\begin{figure}[ht]
\includegraphics[width=1\linewidth]{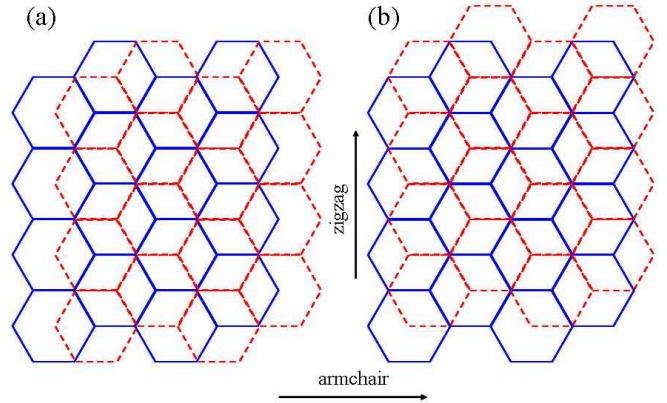}
\caption{ (Color online) Schematic illustration of the two edge alignments we consider in bilayer graphene nanoflakes. (a) $\alpha$-alignment and (b) $\beta$-alignment.}
\label{fig:Figure1}
\end{figure}        

In this article, we use a density functional theory based {\it first principles} method\cite{paolo} and a version of semi-local exchange correlation potential called gradient approximation\cite{perdew} to explore the dependence of the energy gaps of bilayer graphene on geometrical confinements (length as well as width), interlayer magnetism and the external electric fields. 
We assume Bernal (or AB) stacked bilayers with both edge types (armchair and zigzag) and for each edge type,  we consider two edge alignments (Fig.~\ref{fig:Figure1}), denoted as $\alpha$- and $\beta$-alignments. 
Such length confined nanoribbons will be called ``{\it nanoflakes}" throughout this article. In the following discussions, we will refer to gradient approximation as GGA.

We are not aware of any density functional theoretical calculations focussing on interplay of energy gap, magnetism and external electric field in bilayer graphene nanoflakes in the literature.
Only recently, the energy gap of monolayer nanoflakes was studied using a density functional based approach.\cite{nayak} 

Here, we predict interplay of magnetism and geometrical confinement on the energy gap,  
the existence of three distinct classes of gaps in armchair bilayer nanoflakes, and 
external electric field decreasing the energy gaps,
in addition to capturing some of the predictions made for 
monolayer graphene nanoflakes.       
We note that our predicted gap values are useful in optical experiments in which the possibility of exciton formation can be ruled out. The 
presence of bound excitons 
in an absorption spectra, however, can significantly change the intrinsic gaps of the graphene flakes. 

Our paper is organized as follows.  We first summarize the density functional theory (DFT) electronic structure calculations that we have performed, commenting on the motivation for choosing GGA and the different combinations of lengths and widths in section II.  In section III,  we present the results that we have obtained for bilayer nanoflake gaps, focusing most extensively on the interplay between
width and length, 
edge magnetism, and the external electric field between the layers. 
Finally we summarize our results and present our conclusions.

\section{Density Functional Theory Calculations} 

Our electronic structure calculations are performed with plane wave basis sets and ultrasoft pseudopotentials\cite{ultrasoft} implemented in a density functional theory based electronic structure method.\cite{paolo} We use GGA\cite{perdew} as our previous study\cite{sahu1} suggested the inadequacy of the
local density approximation 
(LDA)\cite{perdew1} in capturing the magnetic nature of the ribbons. 
We note that, in the case of $\alpha$-aligned zigzag nanoribbons, the magnetic ground state was found to be sensitive to the nature of the exchange-correlation potential.\cite{sahu1}
Also the choice of GGA enabled us to compare our bilayer nanoflake results with those of the monolayer graphene nanoflakes.\cite{nayak}

We fixed the interlayer separation to 0.335 nm between the nanoflake sheets as GGA is found to highly overestimate the interlayer separation of the graphene sheets.\cite{langreth} This is due to the absence of non-local or van der Waals' interactions
in the LDA, GGA, Perdew-Burke-Ernzerhof (PBE)\cite{burke} 
and the hybrid versions of exchange-correction potential in DFT. 
We placed the nanoflakes in a supercell set-up by introducing vacuum regions of 1 nm along all three superecell directions
to avoid the intercell interactions.
We used 2 $\times$ 2 $\times$ 1 {\bf k}-point mesh in the full Brillouin Zone (BZ) and 
476 eV
kinetic energy cut-offs. The convergence of total energy was carefully tested with larger energy cut-off, vacuum size and the {\bf k}-point mesh. 
Since we are considering finite size systems in a supercell set-up, the energy gaps are computed at the $\Gamma$ point of the irreducible BZ.
To study the behavior of the gaps in the presence of external electric field, we used several values of electric field, perpendicular to the layers, up to a maximum value which is close to the SiO$_{2}$ dielectric breakdown field of 1 V/nm.\cite{maria}

The in-plane $\sigma$-orbitals at the edges were saturated with hydrogen atoms (with the C-H distance chosen to be the C-H bondlength in the CH$_4$ molecule). Since the magnetism arises purely due to the edge geometry and localized carbon $\pi$-orbital, choosing a different C-H bondlength will not alter the interlayer magnetic order. However, we note that different edge functionalizations, other than hydrogens, can alter the electronic structure of the bilayer nanoflakes. Our predictions are therefore most relevant for nanoflakes cut in a hydrogen environment without any significant edge disorder.\cite{footnote}
Moreover, we do not consider edge disorder and the role of different substrates in our calculations.

In order to find the nature of interlayer magnetic order in bilayer nanoflakes, we performed three distinct sets of calculations: a nonmagnetic, a parallel and an antiparallel arrangement of magnetic moments on the carbon atoms between the layers. The intra-edge magnetic order within a layer was assumed to be antiparallel as various theoretical reports suggests.\cite{louie}
To find the magnetic ground state for bilayer armchair and zigzag nanoflakes, we chose a few representative systems. We used both narrow and large width nanoflakes with two edge types and for each edge type, two edge alignments (Fig.~\ref{fig:Figure1}). 

We find that in all cases, the nanoflake with interlayer antiparallel (or antiferromagnetic) magnetic moment arrangement is energetically favorable.\cite{footnote2} This ground state is lower, by about 
0.4 eV,
than the interlayer parallel arrangement of the moments, and the nonmagnetic energy barrier is about 
1 eV
from the ground state of the interlayer antiferromagnetic order. Therefore, all the gap calculations in this article are done with antiferromagnetic order between the layers.\cite{footnote3}

We note that for both edge types, nanoflakes with $\alpha$-alignment is found to be energetically favorable over those with $\beta$-alignments and for the same width and length, $\beta$-aligned edges affect the energy gaps more strongly compared to their non-magnetic counterparts than the $\alpha$-aligned edges. For sake of completeness, we will consider both edge alignments for each edge type in this paper.  

\section{Graphene Bilayer Nanoflake Gaps} 

We now present our results for the width, length, magnetism, and external electric field dependence of the bilayer gaps in nanoflakes with both edge types and two edge alignments. First we discuss the width dependence of both armchair and zigzag nanoflakes with a fixed lengths of  {\it L} $\approx$ 3.2 nm and 3.3 nm respectively. Then we vary the lengths up to 6.1 nm and discuss the dependence of the bilayer gaps on the length. 
It should be noted that an armchair nanoflake with the width/length ratio ({\it W/L}) can equivalently be defined as a zigzag nanoflake with the aspect ratio ({\it L/W}).
We will close this section with the external electric field effects on the gaps with fixed as well as variable lengths.   

\begin{figure}[ht]
\scalebox{0.4}{\includegraphics{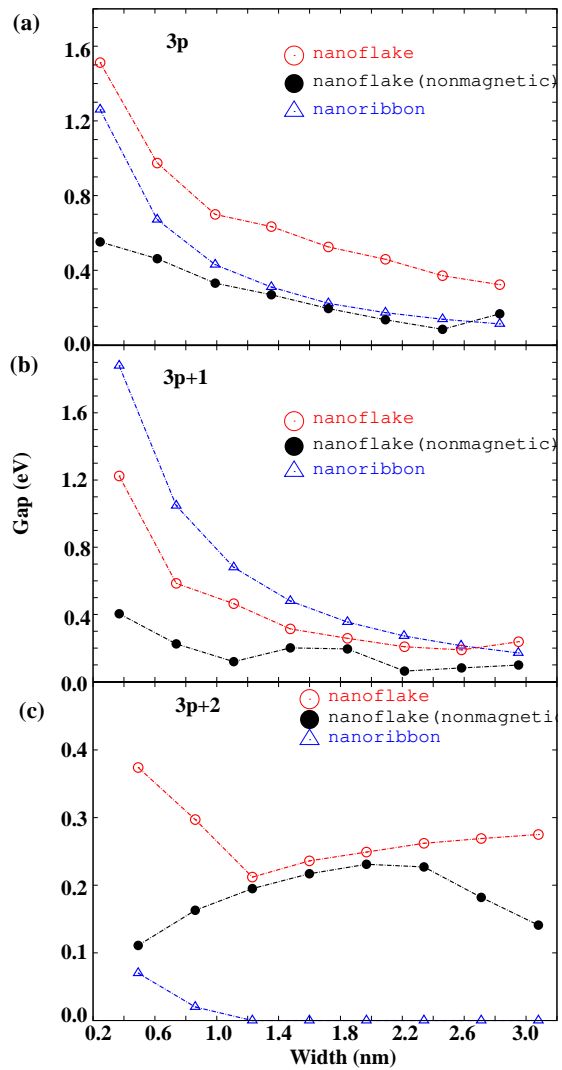}}
\caption{ (Color online) Variation of the energy gap with width of $\alpha$-aligned bilayer armchair nanoflakes for a fixed length {\it L} $\approx$ 3.2 nm (open circles). Three classes of the nanoflakes are clearly seen in (a) 3{\it p}, (b) 3{\it p}+1 and (c) 3{\it p}+2 where {\it p} is an integer specifying the number {\it N} of carbon chains along the width direction. Here {\it p} =1,2,$\cdots$,8, which translate to nanoflakes with widths less than but close to 3 nm.
 The gaps of the magnetic nanoflakes (open circles) are compared with those of the nonmagnetic nanoflakes (solid circles) and three classes of nonmagnetic nanoribbons (open triangles). It should be noted that armchair nanoribbons are all nonmagnetic and all the gaps are obtained using the same semi-local GGA potential.}
\label{fig:fig2}
\end{figure}

\begin{figure}[ht]
\scalebox{0.4}{\includegraphics{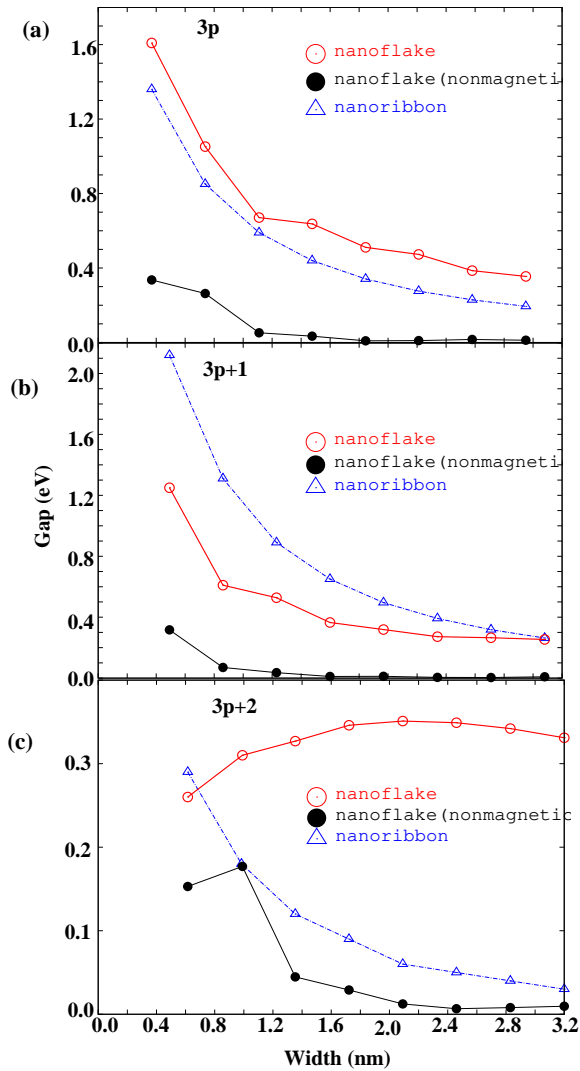}}
\caption{ (Color online) Same as Fig. 2 but for $\beta$-aligned bilayer armchair nanoflakes.}
\label{fig:fig3}
\end{figure}

\subsection{Armchair and zigzag bilayers with fixed length confinement}

In this section, we discuss the width dependence of gaps in 
bilayer graphene nanoflakes
with both edge alignments for a fixed length. 
Figure \ref{fig:fig2} shows variations of the gap versus the nanoflake width for $\alpha$-aligned armchair nanoflakes. For comparison, we also show the gaps for bilayer armchair nanoribbons.  We predict three classes of armchair gaps which we label, according to the widely accepted notations in the literature, by {\it N} = 3{\it p}, 3{\it p}+1 and 3{\it p}+2, where {\it p} is an integer.
As expected on the basis of previous work,\cite{sahu1} the width dependence is quite smooth within the three nanoflake classes, which are distinguished by the number {\it N} of carbon chains across the nanoflake mod 3. However, all classes show a semiconducting behavior. This is in contrast with the class 3{\it p}+2 in armchair nanoribbons which exhibited a tendency towards metallic behavior
whereas the class 3{\it p}+2 in armchair nanoflakes remain semiconducting due to the magnetic zigzag units present along the width direction.

\begin{figure}[ht]
\scalebox{0.4}{\includegraphics{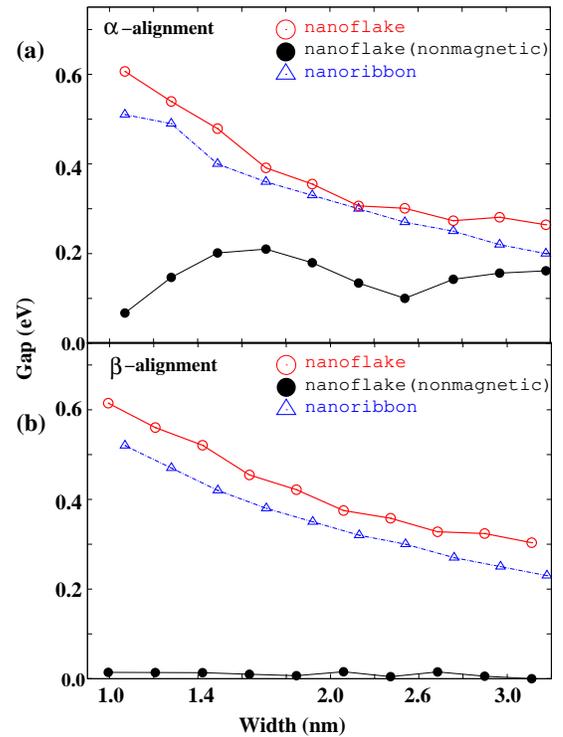}}
\caption{ (Color online) Variations of the energy gap of the (a) $\alpha$-aligned and (b) $\beta$-aligned bilayer zigzag nanoflakes with a fixed length {\it L} $\approx$ 3.3 nm. For comparison, the gap values of nonmagnetic nanoflakes and magnetic nanoribbons are also shown. All the gap calculations are done with the same semi-local GGA potential and for magnetic calculations an interlayer antiferromagnetic order was considered.
}
\label{fig:fig4}
\end{figure}

For the class 3{\it p}, magnetism between the bilayers significantly enhances the gaps compared
not only to nonmagnetic bilayer nanoflakes but also to corresponding nanoribbons,
as seen in Fig.~\ref{fig:fig2}(a).
But this is not immediately apparent for the class 3{\it p}+1 until the width reaches about 3 nm, where the nanoflake gap tends to exceed the nanoribbon gap. To check whether for larger lengths and widths, 
magnetic nanoflake gaps exceed corresponding nanoribbon gaps for 3{\it p}+1,
we considered a longer ribbon ({\it L} $\approx$ 4.2 nm) and a width near 4 nm and found that indeed the bilayer nanoflake gap is larger than the nanoribbon gap (0.27 eV versus 0.12 eV). This indicates an interplay of interlayer magnetism and geometrical confinement in bilayer armchair nanoflakes 
because the number of magnetic zigzag units increases as the width increases, which slows down the decrease of the armchair nanoflake energy gaps compared to nonmagnetic armchair nanoribbon energy gaps.
The behavior of the class 3{\it p}+2 nanoflakes can be understood similarly.
For this class, the length confinement resulted in larger gaps compared to the nanoribbons and the magnetism seems to be contributing strongly in enhancing the gaps for wider nanoflakes, which is opposite to the behavior of wider nanoflakes in two other classes 3{\it p} and 3{p+1}. 
Such increase of class 3{\it p}+2 gaps with the width was also seen in monolayer nanoflakes.\cite{nayak}

DFT again predicts three classes of gaps for $\beta$-aligned nanoflakes (Fig.~\ref{fig:fig3}). Compared to the gaps in $\alpha$-aligned nanoflakes, the gaps are consistently larger for the same width. 
Note that in bilayer nanoribbons\cite{sahu1} the two types of edge alignments provide distinct electronic structures: magnetism in $\beta$-aligned ribbons were more significant than in $\alpha$-aligned ribbons due to a dispersionless nonmagnetic band at the Fermi level spanning $\frac{1}{3}$ of the distance from the edge of the one dimensional BZ although inclusion of magnetic order opened up a gap in the energy spectra of both types.
The two nanoflake classes 3{\it p} and 3{\it p}+1 show similar behavior as in $\alpha$-aligned nanoflakes, again with a hint of order change in class 3{\it p}+1. In class 3{\it p}+2 nanoflakes, the gaps increase with the increase of widths for magnetic nanoflakes while nonmagnetic nanoflakes show behavior similar to the nanoribbons. 
For wider magnetic nanoflakes of {\it W}$>$2.4 nm, however, the energy gaps begin to decrease as width increases, again indicating the complex interplay of magnetism and geometrical confinement. 

Figure \ref{fig:fig4} shows variations of the gap with the widths in $\alpha$- and $\beta$-aligned zigzag nanoflakes. For comparison, gaps of magnetic nanoribbons and nonmagnetic nanoflakes are also shown. 
Interlayer magnetism seems to enhance the gap values compared to the gaps resulting from only geometrical confinement (i.e. nonmagnetic gaps),
but in contrast to the armchair nanoflakes, the energy gaps of zigzag nanoflakes follow the same trend as corresponding zigzag nanoribbons.  
This may be due to the fact that with increasing width in armchair nanoflakes, 
the number of (magnetic) zigzag units increases whereas in zigzag nanoflakes, increasing the width increases the number of (nonmagnetic) armchair units.   

\subsection{Length dependence of the gaps}

\begin{figure}[ht]
\scalebox{0.4}{\includegraphics{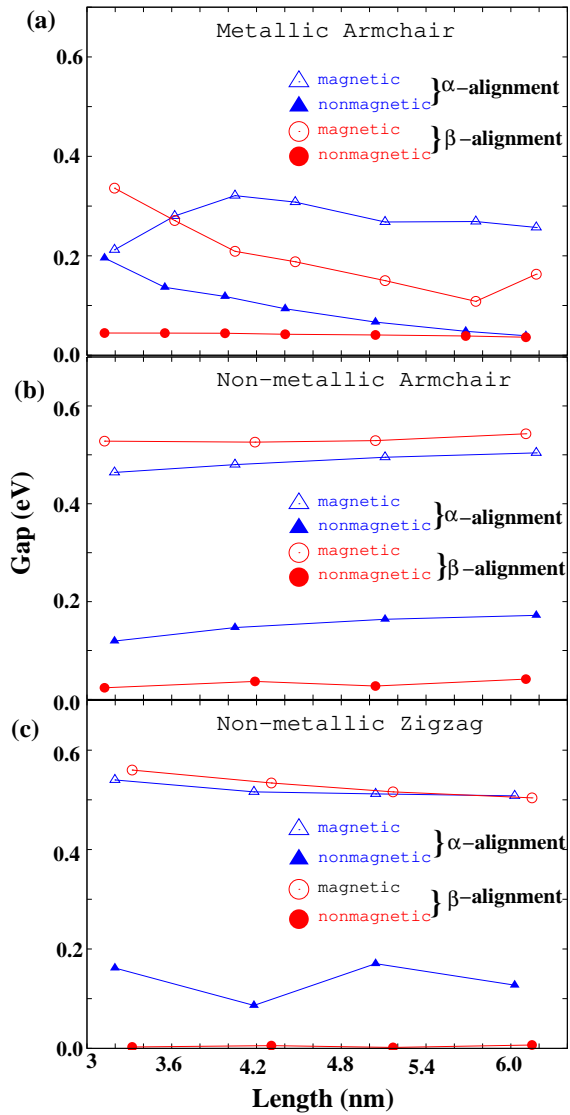}}
\caption{ (Color online) Variation of the armchair and zigzag nanoflake gaps with different lengths are shown. 
In longer nanoflakes, the non-magnetic gaps are shown to be smaller than their magnetic counterpart.  
}
\label{fig:fig5}
\end{figure}

We address the length dependence of the nanoflake gaps in this section. The metallic and non-metallic nanoribbons are expected to respond differently to changes in length as the authors of Ref. 10 show. Therefore, we considered length confined 
metallic (class 3$p$+2) and non-metallic (class 3$p$+1) 
bilayer armchair nanoribbons as well as the non-metallic bilayer zigzag nanoribbons for the study of length dependence of the gaps.  We denote these, respectively, as {\it metallic} and {\it non-metallic} nanoflakes in Fig.~\ref{fig:fig5}. Both edge alignments were used and both magnetic and non-mangetic nanoflakes are considered. We considered representative widths corresponding to {\it N}=10 and 11 ({\it W} = 1.1 nm and 1.3 nm) for armchair nanoflakes and {\it N}=6 ({\it W}=1.3 nm) for zigzag nanoflakes. To keep the computational burden low, we chose few representative lengths ({\it L} $\sim$ 3.2 nm to 6.1 nm).

It is seen from Fig.~\ref{fig:fig5} that non-metallic gaps are weakly dependent on the nanoflake lengths whereas the metallic nanoflake gaps show strong length dependence. 
Moreover, it is clearly seen that interlayer magnetism enhances the {\it intrinsic} gaps in longer nanoflakes.
In Ref. 10, it was suggested that in non-metallic monolayer armchair nanoflakes, the $\pi$-orbitals are strictly localized along the zigzag edges and therefore the gaps are insensitive to the increase in length, whereas for metallic nanoflakes, the orbitals are 
delocalized
throughout the flake and therefore are sensitive to changes in length. We believe that this will also be true for the bilayer nanoflakes except for the fact that 
the bilayer orbitals play a role in the localization/delocalization process and 
the degree of localization will get affected due to interlayer coupling. 
    
\subsection{Electric field effects on the gaps} 

\begin{figure}[ht]
\scalebox{0.4}{\includegraphics{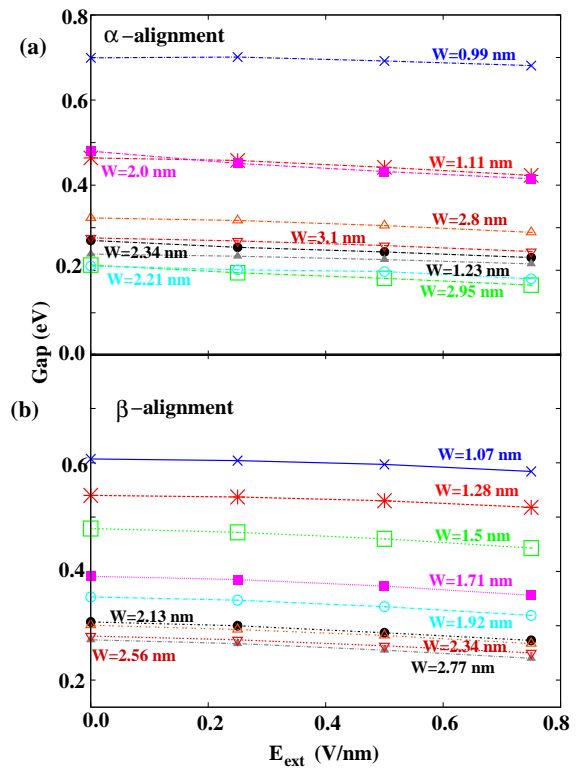}}
\caption{ (Color online) Variation of the energy gap with perpendicular external electric field for bilayer nanoflakes with $\alpha$-alignments for  (a) armchair ({\it L} $\approx$ 3.2 nm) and (b) zigzag nanoflakes ({\it L} $\approx$ 3.3 nm). For each class of armchair nanoflakes, three representative widths specified by the interger {\it p}=3, 6, and 8, are chosen. The gaps are shown at three representative values of the external electric field strength. The maximum electric field strength applied is close to 1 V/nm.}
\label{fig:fig6}
\end{figure}

In this section, we discuss the effect of external electric fields, applied perpendicular to the layers, on the gaps of armchair and zigzag nanoflakes. We considered four different values of the electric fields below the SiO$_{2}$ dielectric breakdown field of 1 V/nm. Both wide and narrow nanoflakes were chosen 
with {\it L} $\approx$ 3.2 nm for armchair and {\it L} $\approx$ 3.3 nm for zigzag nanoflakes. 
The electric field decreases the gap in both armchair and zigzag nanoflakes with $\alpha$-alignments (Fig.~\ref{fig:fig6}), a behavior we predicted for bilayer nanoribbons.\cite{sahu1} A similar behavior is seen for $\beta$-aligned nanoflakes (figures not shown). 

We note that, recently, we predicted the existence of a {\it critical gap} of about 0.2 eV for bilayer nanoribbons below (above) which the electric field has the effect of increasing (decreasing) the gap.\cite{sahu1} We could not verify that such a critical gap can exist for armchair and zigzag nanoflakes because of gap values, all exceeding 0.2 eV, for the chosen widths and lengths.
We did additional calculations of electric field effects on the longer nanoflakes up to {\it L}=6.1 nm to search whether there exists a critical gap but again due to the large gap values in longer nanoflakes, no such critical gap was found.

\section{Summary and Conclusions} 

In summary, we have studied, using a {\it first principles} DFT method, the tunability of bilayer nanoflake gaps as a function of interlayer magnetism, lengths and widths, and external electric fields. Bernal (or AB) type of interlayer stacking with two edge types (armchair and zigzag) and two edge alignments ($\alpha$ and $\beta$) were considered. We identify three distinct classes of armchair gaps and show that interlayer magnetism plays an important role in enhancing the confinement-induced gaps. Length-confined metallic armchair nanoribbon gaps are strongly affected by variations in the length. The energy gap as a function of the applied electric field show decreasing trend for both the edge types and alignments. However, the existence of a critical gap can not be ruled out for nanoflakes with very small intrinsic gaps. We expect that the present results will help stimulate further studies of bilayer gaps in the presence of additional external perturbations such as a substrate and contacts, and motivate experiments to unravel the complicated interplay of magnetism, geometrical confinement and edge type.

\acknowledgments

The authors acknowledge financial support from NRI-SWAN center.
The work done by H. M. has been supported in part by the NIST-CNST/UMD-NanoCenter Cooperative Agreement.
 B. S. acknowledge the allocation of computing time on NSF Teragrid machine {\it Ranger} (TG-DMR080016N) at Texas Advanced Computing Center. 
The authors thank A. H. MacDonald, UT-Austin for helpful discussions. The authors thank M. D. Stiles, J. J. McClelland, K. Gilmore, NIST and L. Colombo, Texas Instruments for their valuable comments.


\begin{thebibliography}{999}
\bibitem{footnote3} We find that the ground state deviate slightly from the ideal interlayer antiferromagnetic order. Therefore, the gap values presented in this article are average of spin-up and spin-down gaps.
\end{thebibliography}
\end{document}